\documentclass[aps,amsmath,amssymb,floatfix,superscriptaddress,nofootinbib,twocolumn,a4paper,prb,showpacs]{revtex4-1}
\usepackage{verbatim}
\usepackage{graphicx}
\usepackage{color}
\usepackage{multirow}
\usepackage{subfig}
\usepackage{float}
\usepackage{caption}
\usepackage{amsxtra}
\usepackage{amscd}
\usepackage{amsthm}
\usepackage{amsfonts}
\usepackage{comment}

\newcommand{\note}[1]{$\color{red}{\bullet}$}

\newcommand{\cM}{\mathcal{M}}

\newcommand{\id}{\mathbf 1}

\newcommand{\ua}{\uparrow}
\newcommand{\da}{\downarrow}

\newcommand{\wf}[1]{SSG}

\begin{document}

\title{Spin-singlet Gaffnian wave function for fractional quantum Hall systems} 

\author{Simon C. Davenport}
\affiliation{Rudolf Peierls Centre for Theoretical Physics, Oxford University, 1 Keble Road, Oxford, OX1 3NP, United Kingdom}

\author{Eddy Ardonne}
\affiliation{%
Nordita, Royal Institute of Technology and Stockholm University,
Roslagstullsbacken 23,
SE-106 91 Stockholm,
Sweden
}
\affiliation{%
Department of Physics, Stockholm University,
AlbaNova University Center, SE-106 91 Stockholm, Sweden
}

\author{Nicolas Regnault}
\affiliation{Department of Physics, Princeton University, Princeton, New Jersey 08544, USA}
\affiliation{Laboratoire Pierre Aigrain, ENS and CNRS, 24 rue Lhomond, F-75005 Paris, France}

\author{Steven H. Simon}
\affiliation{Rudolf Peierls Centre for Theoretical Physics, Oxford University, 1 Keble Road, Oxford, OX1 3NP, United Kingdom}

\date{\today}

\begin{abstract}
We characterize in detail a wave function conceivable in fractional quantum Hall systems where a spin or equivalent degree of freedom is present. This wave function combines the properties of two previously proposed quantum Hall wave functions, namely the non-Abelian spin-singlet state and the nonunitary Gaffnian wave function. This is a spin-singlet generalization of the spin-polarized Gaffnian, which we call the ``spin-singlet Gaffnian'' (SSG). In this paper we present evidence demonstrating that the SSG corresponds to the ground state of a certain local Hamiltonian, which we explicitly construct, and, further, we provide a relatively simple analytic expression for the unique ground-state wave functions, which we define as the zero energy eigenstates of that local Hamiltonian. In addition, we have determined a certain nonunitary, rational conformal field theory which provides an underlying description of the SSG and we thus conclude that the SSG is ungapped in the thermodynamic limit. In order to verify our construction, we implement two recently proposed techniques for the analysis of fractional quantum Hall trial states: The ``spin dressed squeezing algorithm'', and the ``generalized Pauli principle''. 
\end{abstract}

\maketitle


\section{Introduction}

The study of the fractional quantum Hall effect (FQHE) of electrons in semi-conductor materials \cite{prangebook} and, potentially, of bosons in rotating Bose gasses \cite{cooper2008} or in artificial gauge fields \cite{cooper2011} has been a fertile ground for the conception of topological phases of matter, which exhibit exotic and exciting theoretical properties. \cite{nayak2008}  In this paper we shall discuss the construction of a FQHE wave function exhibiting a ``multicomponent'' degree of freedom such as a spin, \cite{du1995} valley, \cite{bishop2007,dean2011} isospin, \cite{eng2007} layer, \cite{eisensteinbook} or subband. \cite{hormozi2012,palmer2006} The wave function describes the low-energy behaviour of a system possessing a vanishing energy gap to quasihole excitations in the thermodynamic limit, and is thus not itself a candidate to describe a topological phase. Nevertheless the study of this wave function is an interesting exploration that enhances the understanding of the construction of ``multicomponent'' FQHE wave functions in general.

An ideal theoretical description of a FQHE state comprises at least the following three ingredients: A local Hamiltonian which describes the ground state and excitation spectrum of the Hilbert space; relatively simple analytic wave functions which are the highest density zero energy states corresponding to the Hilbert space of that Hamiltonian; and a (rational) two-dimensional (2D) conformal field theory (CFT) which generates these wave functions. If a plasma analogy is available, then a gapped state is associated with the analogous plasma being in a screening phase. \cite{laughlin1983a} In general, the quantum Hall Hilbert space is built from a basis of monomials in the complex particle coordinates $z_i$.  In certain instances---the Laughlin series, the Read--Rezayi series, the Gaffnian and Haffnian wave functions, the Halperin wave functions and the non-Abelian spin-singlet (NASS) states \cite{laughlin1983a,read1999, simon2007b,green2002,halperin1983,ardonne1999,ardonne2001,ardonne2002} ---it has been possible to i) determine simple analytical expressions for the polynomial wave functions, and ii) construct a  local Hamiltonian whose zero energy eigenstates are in one-to-one correspondence with those polynomial wave functions. A key feature in each of these special cases is that the wave functions are uniquely defined by their vanishing properties.

Changing wave functions describing a FQHE state in a seemingly innocuous way, may in fact have severe consequences. The wave function of the Moore--Read state for spinless bosons vanishes quadratically when three of the constituent bosons are coincident. The Gaffnian wave function is obtained by changing this behaviour, such that the wave function vanishes as a third power, when three constituent particles are coincident. \cite{simon2007b} This simple change, however, results in a nonunitary, compressible wave function, which does not describe a topological phase. Nevertheless nonunitary wave functions such as the Gaffnian are still of interest as they are thought to correspond to critical points between other unitary, incompressible topological phases. This scenario is well understood in the case of the Haldane--Rezayi\cite{haldane1988} wave function, which describes the phase transition between a $d$-wave spin-singlet phase and a strongly paired state. \cite{read2000,green2002} Similar scenarios have been suggested for the Gaffnian.\cite{simon2007b} It is noteworthy that the Gaffnian wave function has large overlap with an incompressible composite fermion state thus suggesting that the Gaffnian is a critical point between the composite fermion phase and some other phase. \cite{jain1989,simon2007b,regnault2008,regnault2009} 

In this paper, we present evidence for a quantum Hall wave function, which inherits properties from both the NASS state---a unitary, spin-singlet quantum Hall state---and the nonunitary Gaffnian. Such a wave function has also been considered in Ref.~\onlinecite{estienne2011}. Most notably, we argue that the wave function can be written as a special polynomial and that it also corresponds to the highest density zero energy state of a certain local Hamiltonian. We christen it the ``spin-singlet Gaffnian'' (\wf{}). (Throughout this work, we shall describe how the wave function corresponds to a ``spin'' degree of freedom, however, note that all of what follows applies equally well to any other type of multicomponent degree of freedom such as valley or layer index.) 

\subsection*{Statement of results}

Our first key result is a model local Hamiltonian describing the \wf{} wave function. In Sec.~\ref{secPseudopotential} we shall explain how it can be written in terms of generalized Haldane pseudopotentials. \cite{haldane1983,simon2007a,davenport2012a} In Secs.~\ref{secSqueezing} and \ref{secPauli}, we give an overview of how both the squeezing algorithm \cite{bernevig2008b,ardonne2011} and the generalized Pauli principle for spinful states \cite{estienne2011}  apply to the \wf{} wave function.  In Sec.~\ref{secNumerics} we shall present the results of numerical exact diagonalization of the \wf{} Hamiltonian. We have determined that the counting of zero energy eigenstates in the quasihole spectrum and also the entanglement spectrum \cite{li2008,sterdyniak2011} matches the counting predicted by the squeezing algorithm. The counting of zero energy eigenstates is also found to be identical to the result generated by the spinful version of the generalized Pauli principle. 

Our second key result is an analytic form for the ground-state wave function. The \wf{} wave functions are constructed from conformal blocks in a CFT associated with the semidirect product of nonunitary minimal models expressible as $\mathcal{M}(3,5) \ltimes \mathcal{M}(5,7)$ (see also Ref. \onlinecite{estienne2011}). We shall discuss the CFT in Sec.~\ref{secCft}. In a Bosonic incarnation, the \wf{} ground state occurs at filling factor $\nu=4/5$ (and on the sphere, it has a shift $\delta=3$). The proposed analytic construction for the \wf{} is consistent with the vanishing properties required by the CFT and the ground state of the Hilbert space generated by squeezing.  We have determined that our proposed ground-state wave function is the unique, highest density zero energy ground state of our model Hamiltonian at $\nu=4/5$ and $\delta=3$. We shall discuss an explicit form of the ground-state wave function in Sec.~\ref{secwave function}.


\section{Model Hamiltonian and the Characterization of its Zero Modes}
\label{secMethods}

In this section we shall describe in detail three independent methods by which we are able to study the \wf{} wave function. The cornerstone of our argument lies with the proposal of a model local Hamiltonian. First, we shall explain how this local Hamiltonian is constructed and then we shall present two alternative methods to count the number of zero energy modes of the proposed Hamiltonian. The first method employs the spin-dressed squeezing algorithm (which can also be used to obtain the wave functions) and the second method employs the generalized Pauli principle. In Sec.~\ref{secNumerics} we shall present numerical evidence demonstrating that the zero modes generated by diagonalizing the local Hamiltonian presented here are in precise agreement with the zero modes generated by the squeezing algorithm and the counting obtained from the generalized Pauli principle. 

\subsection{Pseudopotential construction of Hamiltonian}
\label{secPseudopotential}

In prior investigations, most notably for the Laughlin and Moore--Read wave functions, it has been determined that trial quantum Hall wave functions correspond to unique, zero energy ground states of certain model Hamiltonians. \cite{haldane1983,greiter1992} Often, these model Hamiltonians are most simply expressible in terms of Haldane pseudopotentials \cite{haldane1983} and their generalizations. \cite{simon2007a,davenport2012a} Given a system with a certain $M$-body interaction potential $V(z_1,\ldots, z_M)$, the action of a pseudopotential is to project out a particular component of that interaction.  Those components are labelled by a convenient set of quantum numbers, which describe all possible few-particle interactions. The vector space of few-particle interactions is spanned by specifying both the relative angular momentum $L$ and, if we consider a spin degree of freedom as well, the spin quantum number $S$ also. (Note that in general it is necessary to further distinguish between distinct interaction components with the same $L,S$. In other words, there exists in general a subvector space of dimension $d_{L,S}$ for each $L$ and $S$ sector. \cite{davenport2012a,simon2010} In this paper, however, we shall only be concerned with subspaces of dimension $d_{L,S}$=1 or 0, and so, for clarity, we omit any additional notation.) Pseudopotentials specify projection operators in the Hamiltonian and are thus labelled by two distinct sets of such quantum numbers. The pseudopotential $V^M_{L,S ; L' S' }$ is expressed in terms of the vector space $\left|  L, S \right \rangle$ and the $M$-body interaction potential $V(z_1,\ldots, z_M)$ as
\begin{equation}
\label{eqMulltiParticlePseudopotential}
V^M_{L,S ; L' S' } = \left\langle L, S \right |      V(z_1,\ldots, z_M)        \left|  L', S' \right \rangle.
\end{equation}
The general form of the Hamiltonian is then given by
\begin{equation}
\label{eqGeneralHamiltonian}
H=\sum_{L, S, L' S'}    \left|  L, S \right \rangle  V^M_{L,S ; L'S' } \left\langle L', S' \right |.
\end{equation}
Such a basis of pseudopotentials is particularly convenient when the interaction potential is rotationally and/or spin rotationally invariant, since in that case the pseudopotentials are diagonal in $L$, $S$, or both. 

At our convenience, we can pick certain special many-body interactions (such as $\delta$-function-type interactions) for which only a small set of pseudopotentials remain nonzero. The impact of specifying a positive value of a given pseudopotential in a model Hamiltonian is to assign energy to the corresponding component of the interaction, therefore, if such a component is present in a given trial wave function, then that wave function will not be a zero energy state of our Hamiltonian. Conversely, if a component is not present in a given trial wave function, then we can include the corresponding pseudopotential in the Hamiltonian without introducing an extra energy cost. In this way we can tailor the Hamiltonian to correspond to the desired properties of a given trial ground-state wave function (these properties might come, for instance, from a CFT description of the state; see Sec. \ref{secCft}).

To give an example, briefly, it is known that the Moore--Read wave function for bosons at $\nu=1$ is the unique, highest density zero energy ground state of a spin-polarized three-body contact interaction. \cite{greiter1992} The space of pseudopotentials for spin-polarized three-body interactions is spanned by a relative angular momentum $L$. $S$ takes only its maximal value for a three-body interaction, $S=3/2$. \cite{simon2007a} (Note a slight abuse of notation here: We are denoting bosons as if they are spin-1/2 objects, e.g., three bosons can have $S=3/2$ and $S=1/2$ interaction channels. In actual fact we are describing systems such as cold atomic gasses for which bosons can be engineered to have internal two-state degrees of freedom, and we have simply mapped that two-state degree of freedom onto a spin degree of freedom.) In the language of pseudopotentials,  the Moore--Read state corresponds to $ V^3_{0,3/2; 0,3/2}$ being positive and all other pseudopotentials being zero and, in addition, the corresponding Hilbert space is restricted to only spin-polarized sectors. Along similar lines, the spin-polarized Gaffnian wave function for bosons corresponds to  $ V^3_{0 ,3/2; 0,3/2}$ and  $ V^3_{2,3/2 ;2,3/2}$ being positive and all other pseudopotentials being zero, and again  the Hilbert space is restricted to spin-polarized sectors (note that for $L=1$, $S=3/2$, and $M=3$, we have a 0-dimensional vector space of wave functions, so no corresponding $L=1$, $S=3/2$, and $M=3$ pseudopotential can occur). In these examples we observe that the pseudopotentials are in fact diagonal in the $L,S$ sectors. This is a direct consequence of the interaction being, respectively, rotationally and spin-rotationally invariant. This feature will remain present in our application of pseudopotentials to the \wf{} wave function, and so at this point we shall drop the repeated indices and denote diagonal pseudopotentials by $V^M_{L,S} \equiv V^M_{L,S ; L S } $.

The Bosonic NASS state, like the Moore--Read state, corresponds to a spin-polarized three-body contact interaction, however the Hilbert space now includes additional spin sectors (not just the spin-polarized sector).  In the language of pseudopotentials,  the Bosonic NASS state is the highest density zero energy ground state of a Hamiltonian with the positive, diagonal pseudopotential $ V^3_{0,3/2}$. Motivated by the generalization of the Moore--Read Hamiltonian to the Gaffnian Hamiltonian, our proposal for the \wf{} Hamiltonian is to keep $ V^3_{0,3/2}$, $ V^3_{2,3/2}$, and $ V^3_{1,1/2}$ positive. The proposed Hamiltonian for the \wf{} wave function is thus expressed as
\begin{align}
\label{eqHamiltonian}
H_{\mbox{\wf{}}} &= \left|  0, 3/2 \right \rangle  V^3_{0,3/2}  \left\langle 0, 3/2 \right | \nonumber \\ &+ \left|  2, 3/2 \right \rangle  V^3_{2,3/2}  \left\langle 2, 3/2 \right | \nonumber \\&+ \left|  1, 1/2 \right \rangle  V^3_{1,1/2}  \left\langle 1,  1/2 \right |.
\end{align}

From this pseudopotential Hamiltonian, we can already infer some properties of its zero energy eigenstates. Because the three-body interaction in the $S=3/2$ channel is identical to the Hamiltonian generating the (polarized) Gaffnian, the zero energy eigenstates will vanish as at least a third power when three particles with the same spin are brought to the same point. In addition, when three particles have overall spin $S=1/2$, the wave functions vanish at least quadratically. These are indeed the vanishing properties consistent with the CFT description of the \wf{}, as we shall describe in Sec.~\ref{secCft}. It is worth stating here that the \wf{} wave function likely also corresponds to the ground state of other, more complicated, local Hamiltonians involving $M$-body terms with $M>3$, following the line of reasoning discussed in Ref.~\onlinecite{simon2007a}.

\subsection{Spin-dressed squeezing algorithm}
\label{secSqueezing}

There is another way to characterize the spin-singlet Gaffnian, apart from by means of the Hamiltonian we introduced above. This method is the so-called ``squeezing method''. The idea underlying the method is that for many model states, the wave functions have a large number of zero coefficients if expressed in terms of the space of all possible monomials. These zeros are closely related to the vanishing properties of the wave functions. The method was pioneered in Ref.~ \onlinecite{bernevig2008b}. The spinful case, which we will employ here, was described in great detail in Ref.~\onlinecite{ardonne2011}. 

We consider quantum Hall states on the sphere, \cite{haldane1983} in the presence of $N_\Phi$ flux quanta. In the lowest Landau level, this construction gives rise to $N_\Phi+1$ orbitals, whose occupation numbers will be denoted by $(n_0,n_1,\ldots,n_{N_\Phi})$. These orbitals have angular momentum $(N_\Phi/2, N_\Phi/2-1,\ldots,-N_\Phi/2)$. We will use the orbital occupation numbers to label the states in the Hilbert space. Because total angular momentum is a good quantum number, we can split the Hilbert space into sectors with different values of total $L_z$, and we will always consider the number of particles $N$ to be fixed. For spinless fermions, the Pauli principle specifies that the occupation of any orbital $n_i$ can be 0 or 1, while for bosons, there is no constraint.

To obtain all states in a particular $L_z$ sector, we divide the $N$ particles over the orbitals, such that one obtains the correct value of total $L_z$, and such that the particles are ``desqueezed'' as much as possible. This means that for an even number of bosons in the $L_z=0$ sector, one considers the occupation $(N/2,0,\ldots,0,N/2)$. To obtain the other states in this sector, one generates all possible pairwise rearrangements of particles occupying orbitals, keeping the total $L_z$ fixed. In the general case, this entails $(\ldots,n_{i},n_{i+1},\ldots,n_{j-1},n_{j},\ldots)$ transforming into   $(\ldots,n_{i}-1,n_{i+1}+1,\ldots,n_{j-1}+1,n_{j}-1,\ldots)$, where all the other occupation numbers remain unchanged. In this way, one obtains the {\em full} Hilbert space in each total $L_z$ sector.

As we alluded to above, the wave functions of many model states have a large number of zero components, if expressed in the Hilbert space described above. For instance, if one considers the $\nu=1/2$ Bosonic Laughlin state, it suffices to construct a {\em reduced} Hilbert space, by starting to squeeze from the following so-called {\em root configuration}, $(1,0,1,0,\ldots,0,1,0,1)$, instead of the completely desqueezed configuration $(N/2,0,\ldots,0,N/2)$. All states in the {\em full} Hilbert space which do not appear in the {\em reduced} Hilbert space have zero coefficient in the Laughlin state. Moreover, it turns out that there is only one $L=0$ state one can construct in the {\em reduced} Hilbert space. Thus, to obtain the Laughlin state for a certain number of particles, one constructs the reduced Hilbert space from the root configuration $(1,0,1,0,\ldots,0,1,0,1)$, and demands that $L^{+}$ on a general state in this reduced Hilbert space gives zero. This procedure gives a set of equations for the coefficients, whose solution gives the Laughlin state, expressed in the monomial basis on the sphere. In a similar way, many other model states can be obtained. For instance, to obtain the level-$k$ Read-Rezayi states \cite{read1999}, one only has to change the root configuration to $(k,0,k,\ldots,k,0,k)$. Interestingly, the root configurations correspond to the ``thin-torus'' limit of the states. \cite{bergholtz2005,bergholtz2006,seidel2006}

To describe spinful wave functions, the squeezing method was augmented in Ref.~\onlinecite{ardonne2011}. The idea is to start with a {\rm root configuration}, for which one at first ignores the spin degrees of freedom. To construct the reduced Hilbert space, one creates a set of orbital occupations, by squeezing in all possible ways. Finally, one assigns spin to all the particles, in all possible ways. We will be interested in spin-singlet states, for which total $S$ is a good quantum number. So to construct the ground state, we work in the total $S_z=0$ sector. After distributing the spin in all possible ways for each configuration, we have constructed the reduced Hilbert space.  The spinful orbital occupations are now denoted by $(n_{0,\uparrow},n_{0,\downarrow},n_{1,\uparrow},n_{1,\downarrow},\ldots,n_{N_{\Phi},\uparrow},n_{N_{\Phi},\downarrow})$ . 

We still have to specify how to obtain the correct wave function in the reduced Hilbert space. In general, we will be interested in ground-state wave functions which have $L=S=0$, so to obtain those, we work in the $L_z=S_z=0$ subsectors, and demand that the action of $L^{+}$ and $S^{+}$ on the state gives zero. It turns out that this is in general not enough to completely specify the wave functions. In addition, one has to demand that certain states in the reduced Hilbert space have zero coefficient. The SSG is an example where this happens, as we shall presently describe.

Let us now discuss how the squeezing algorithm applies to the construction of the \wf{} Hilbert space. We have already seen, in the construction of the Hamiltonian in Eq.~(\ref{eqHamiltonian}), that the zero energy eigenstates of the \wf{} wave function vanish as at least a third power when three particles with the same spin are brought to the same point and, in addition, when three particles have overall spin $S=1/2$, the wave functions vanish at least quadratically. These observations motivate the use of the root configurations based on the pattern $(2,0,2,0,0,2,0,2,0,0,\ldots,0,0,2,0,2) $, because like the Moore--Read state, the resulting wave function will vanish at least quadratically when three particles are at the same location. These root configurations have the property that two neighbouring orbitals can maximally be occupied by two particles, while five consecutive orbitals can be occupied by at most four particles. We note in advance that these root configurations satisfy the spinful generalized Pauli principle, which we discuss in the next subsection.

To construct the ground-state wave function, we start with the root configuration described above, and then construct the reduced Hilbert space. To obtain the ground state, we then demand that both $L^{+}$ and $S^{+}$ act to give zero on all states in the reduced Hilbert. This construction, however, only guarantees that the obtained wave function vanishes as a second power when three particles of the same spin are coincident; but we really wanted the wave function to vanish as a third power in that case. We therefore enforce an additional constraint: That all basis states in the reduced Hilbert space have zero coefficient, on the condition that either one has for spin-up particles $n_{0,\uparrow} = 2$ and $n_{2,\uparrow} > 0$, or similarly, for spin-down particles, one has $n_{0,\downarrow} = 2$ and $n_{2,\downarrow} > 0$. With that additional constraint, the wave function now vanishes as a third power when three particles are coincident with total $S=3/2$.

\subsection{Spinful generalized Pauli principle}
\label{secPauli}

A further method to characterize the zero energy  space of the SSG Hamiltonian is to use the generalized Pauli principle for quantum Hall wave functions. \cite{bernevig2008a} The generalized Pauli principle  is most readily expressed in terms of partitions (or occupied orbitals).  A partition ${\lambda}$ is defined to be an ordered set of $N$ integers, \{$\lambda_1,\ldots,\lambda_N$\}, where $\lambda_i\geq \lambda_{i+1}$.

To describe the Hilbert space of spinless particles in flux $N_\Phi$, we restrict the integers to the set $\lambda_i \in \{0,1,\ldots,N_{\Phi} \}$. We then have the following relation between the orbital occupation numbers $n_{0},n_{1},\ldots,n_{N_\Phi}$, which were introduced in the previous subsection, and the
$\lambda_i$ forming the partition $\lambda$: Namely, $n_{N_\Phi}$ is the number of $i$ such that $\lambda_{i} = N_\phi$ etc.  Thus in general,
$n_{j}$ is the number of $i$ such that $\lambda_i=j$.

To characterize the quasiholes of the $(k,r)$ clustered states\cite{bernevig2008c} we introduce the notion of $(k,r)$-admissible partitions, which are partitions obeying the following condition for all $i$:
\begin{equation}
 \lambda_i - \lambda_{i+k} \geq r.  \label{PauliSpinless}
\end{equation}

For given fixed values of $N$ and $N_\Phi$, one identifies the number of admissible partitions with the number of quasihole states in the corresponding spectrum. For instance, the $r=2$ series corresponds to the level-$k$ Read--Rezayi states, the Laughlin state $(k=1)$, and the Moore--Read state $(k=2)$ being the simplest cases. The Gaffnian wave function is associated with the $(k=2,r=3)$ generalized Pauli principle.

In order to generalize this method to the spinful case, we follow the argument presented in Ref.~\onlinecite{estienne2011}. First, the partition is replaced by a spinful partition that mixes momentum and spin. A spinful partition $(\lambda,\sigma)$ is specified by $N$ integers, \{$\lambda_1,\ldots,\lambda_N$\}, and now in addition, a set of spin indices  $\{\sigma_1, \ldots \sigma_N\}$, where $\sigma_i \in \{-1, 1\} $. If either the condition
$\lambda_i>\lambda_{i+1}$ or $\lambda_i=\lambda_{i+1}$ and $\sigma_i \geq \sigma_{i+1}$ holds, we say that $(\lambda,\sigma)$ constitutes a spinful partition.

The spinful generalization of the $(k,r)$ admissible partitions is given by the following conditions, which have to hold for all $i$:
\begin{eqnarray}
&& \lambda_i - \lambda_{i+k} \geq r \nonumber,\\
&\mathrm{~or~}& \lambda_i - \lambda_{i+k} = r - 1 \mathrm{~and~} \sigma_i < \sigma_{i+k}\label{Paulispinful}.
\end{eqnarray}

In Ref.~\onlinecite{estienne2011} it is shown that the densest Bosonic state that leads to a spinful $(k,r)$ admissible possible corresponds to a filling factor $\nu = 2k/(2r-1)$ and a shift $\delta = r$. Note that on the sphere, the relation between the filling factor $\nu$ and shift $\delta$ is
\[
\nu=\frac{N}{N_\Phi+\delta}.
\]

It has been proposed \cite{estienne2011} that nonsymmetric Jack polynomials can represent spin-singlet states that are the spin generalization of the spinless clustered states. In that case, the root partition is chosen to be a spinful $(k,r)$ admissible partition. The number of quasihole states can be obtained in a similar way as the spinless case, counting the number of admissible partitions. Indeed, these properties have been checked for the Halperin spin-singlet states $(k=1,r)$ and the NASS state $(k=2,r=2)$. In our application of the spinful generalized Pauli principle to the \wf{}, we employ the $(k=2,r=3)$ Pauli principle (as for the spin-polarized Gaffnian), but now using the spinful partitions. 

\section{Numerical results}
\label{secNumerics}

In this section we shall present numerical evidence demonstrating that the zero energy states of the local Hamiltonian presented in Sec.~\ref{secMethods}, Eq. (\ref{eqHamiltonian}), are correctly reproduced by both the squeezing algorithm and by the generalized Pauli principle which were also described in Sec.~\ref{secMethods}. We shall present evidence derived from both the quasihole spectrum and particle entanglement spectrum.

\subsection{Quasihole spectrum}
\label{secSpectrum}

\begin{figure}[t,b]
\includegraphics[width=0.5\textwidth]{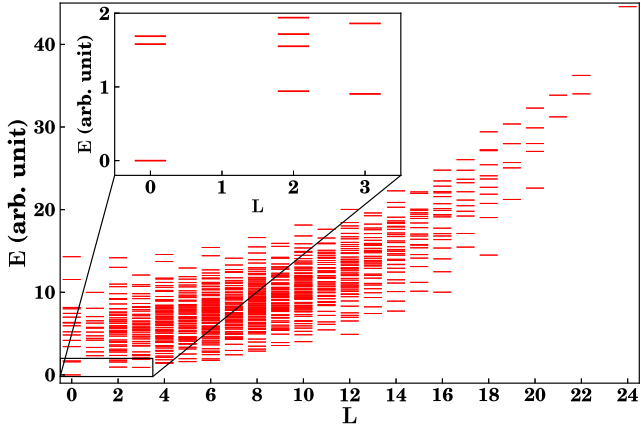}
\caption{(Colour online) The $S=0$ sector of the quasihole spectrum of the Hamiltonian $H_{\mbox{\wf{}}} $, defined in Eq.~(\ref{eqHamiltonian}), for $N=8$ and $N_\Phi=7$, obtained by numerical exact diagonalization in the sphere geometry. The inset zooms in on the bottom left corner of the spectrum and shows the unique zero energy state located in the $L=0$, $S=0$ sector.}
\label{figSpectrum}
\end{figure}

The quasihole spectrum is determined by numerical exact diagonalization of the Hamiltonian for finite-sized systems in the sphere geometry. This is done for a variety of system sizes, $N$, and for a variety of fluxes $N_\Phi$. Eigenstates are labelled by the quantum numbers $L_z$ and $S_z$ and fall into $(L,S)$ multiplets. (Note that these quantum numbers are completely separate from the $L$ and $S$ describing the pseudopotentials unless one has only $M$ particles with an $M$-body interaction.) Given constraints on the dimension of the spinful Hilbert space, we were able to study systems of up to $N=12$ and $N_\Phi=12$. An example of such a spectrum is plotted in Fig.~\ref{figSpectrum}. 

For comparison, we generated the same Hilbert space via the spin-dressed squeezing and generalized Pauli algorithms. For the spin-dressed squeezing algorithm we started from a root partition $(2,0,2,0,0)$, e.g.,  $(2,0,2,0,0,2,0,2,0,0,2,0,2)$ for $N=12$, and we applied the procedure described in Sec.~\ref{secSqueezing} to generate the Hilbert space. The ground-state wave functions, corresponding to the densest root configurations with $L_z=0$, occur at filling $\nu=4/5$ and shift $\delta=3$. For the generalized Pauli algorithm, we specified a spinful admissible partition $(k=2,r=3)$.

Our key observations are as follows: First, in the spectrum of the Hamiltonian there is a unique zero energy state occurring only in the $L=0$, $S=0$ sector for $N=4$ at $N_\Phi=2$, for $N=8$ at $N_\Phi=7$ (see Fig.~\ref{figSpectrum}) and for $N=12$ at $N_\Phi=12$, which all correspond to $\nu=4/5$ and $\delta=3$, consistent with both the squeezing and generalized Pauli approaches; second, for the quasihole spectrum generated for $N$ between 3 and 12 and for $N_\Phi$ up to 20, we have checked that the counting of zero energy states in each $L_z$ and $S_z$ sector precisely matches the counting predicted by both the spin-dressed squeezing algorithm and the generalized Pauli principle.

We have further determined for $N=4$ and $N=8$ that the ground-state monomial expansion of the Hilbert space generated by exact diagonalization of Eq.~(\ref{eqHamiltonian}) precisely matches the ground-state monomial expansion generated by the squeezing algorithm as described in Sec.~\ref{secSqueezing}. In addition we have found that the ground-state monomial expansion also precisely matches that of the proposed analytic form of the ground-state \wf{} wave function (we shall discuss the analytical form of the ground-state wave function in Sec.~\ref{secwave function}).

\subsection{Particle entanglement spectrum}
\label{secPES}

The so-called particle entanglement spectrum (PES) is determined from the reduced density matrix of a subsystem that results from the partition of the whole system into two (or more) parts $A$ and $B$. \cite{li2008, sterdyniak2011} During this partitioning we also keep the overall geometry unchanged. The reduced density matrix $\rho_A$ is given in terms of the full density matrix $\rho=\sum \left | \psi \right \rangle \left \langle \psi \right |$ by tracing out the $N_B$ particles in the $B$ partition: $\rho_A = \mbox{Tr}_B (\rho) $. The PES arises from diagonalizing $\rho_A$ and then classifying the resulting eigenstates according to the symmetries of the problem, in our case by $L_A$ and $S_A$. 

For model states for topological phases such as the Laughlin and Moore--Read states, it has been observed that there is a characteristic PES: The number of nonzero eigenvalues for $\rho_A$ is identical to the number of quasihole states for a system with identical geometry but only $N_A$ particles.\cite{ li2008, sterdyniak2011} The number is usually exponentially lower than the dimension of $\rho_A$, or equivalently, there is an infinite so-called ``entanglement gap'' to the remaining eigenvalues. \cite{li2008,zozuly2009,thomale2010,hermanns2011}

\begin{figure}[t,b]
\includegraphics[width=0.5\textwidth]{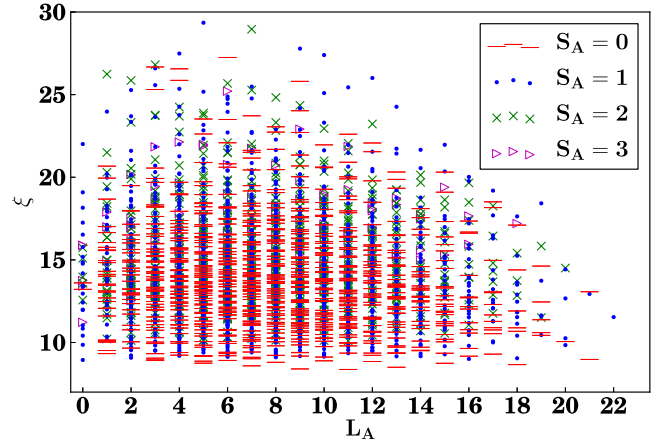}
\caption{(Colour online) The particle entanglement spectrum (PES) for $N=12$, $N_\Phi=12$ and $N_A=6$, obtained by numerical exact diagonalization in the sphere geometry. The negative log of the eigenvalues, $\xi$, of the reduced density matrix $\rho_A$ (defined in the text) are plotted.}
\label{figPesSpectrum}
\end{figure}

For the \wf{} wave function, our key observations are as follows: We find evidence of an infinite entanglement gap for $N=4$ at $N_\Phi=2$, for $N=8$ at $N_\Phi=7$ and for $N=12$ at $N_\Phi=12$ (see Fig.~\ref{figPesSpectrum}), which we associate with our model ground-state wave function for the \wf{}; and we find that the counting predicted for the quasihole excitations up to $N=12$, obtained by considering the PES for all possible partitions of the system into $A$ and $B$ subsystems, precisely matches the counting predicted by the generalized Pauli principle in each $S$ and $L$ sector. 

With the evidence presented in this section in mind, we conclude that the proposed \wf{} Hamiltonian in Eq.~(\ref{eqHamiltonian}) generates the correct zero energy Hilbert space for the \wf{} wave function, at least for system sizes up to $N=12$. Based on this evidence, we also expect our conclusion to hold for any other system size. Thus we have a Hamiltonian which describes the ground state and excitation spectrum of the Hilbert space.

An important question, left unanswered by our numerical calculations here, is what is the size of the gap for quasiparticle excitations in the thermodynamic limit? For finite-sized systems, looking at Fig.~\ref{figSpectrum}, the gap is clearly finite (as it is in the Gaffnian case). There is good reason, however, to argue that the gap will in fact vanish in the thermodynamic limit, and hence the \wf{} will be gapless and the wave function therefore compressible (see Ref.~\onlinecite{read2009}). That argument is based on the fact that the CFT corresponding to the \wf{} is nonunitary, as we shall presently discuss. 

\section{Conformal Field Theory}
\label{secCft}

Let us now discuss the guiding principles of the derivation of the CFT for the \wf{} wave function (see also related considerations in Ref. \onlinecite{estienne2011}). We shall expand further on the discussion in the Appendix. For an introduction on CFT, we refer to the book, Ref. ~\onlinecite{byb}, and the seminal paper by Belavin, Polyakov, and Zamolodchikov (BPZ), Ref~\onlinecite{belavin1984}.

\subsection{Coset constructions for minimal models}

The minimal models introduced by Belavin, Polyakov, and Zamolodchikov can be written in terms of coset models of the $su(2)_k$ WZW model. \cite{goddard1985} In the unitary case, one has
\begin{equation}
\cM (k+1,k+2) = \frac{su(2)_1\times su(2)_{k-1}}{su(2)_k} \ .
\end{equation}
For $k=2$, this gives the Ising CFT, $\cM (3,4)$. For integer $k$, the coset theories are unitary, a property which is inherited from the $su(2)_k$ WZW model, which is unitary for $k$ integer.

General minimal models are labeled $\cM(p',p)$ for arbitrary (non-negative) co-prime integers $(p',p)$. The minimal model is unitary only if $| p - p'| = 1$, otherwise it is nonunitary. Nevertheless, there exists a coset description of nonunitary minimal models in terms of fractional level WZW models. \cite{mathieu1990,mathieu1992} In particular, the parameter $k$ is given by $k=\frac{3p'-2p}{p-p'}$. The nonunitary minimal model $\cM(3,5)$, featuring in the CFT description of the Gaffnian wave function, corresponds to $k=-1/2$, while the Yang-Lee model $\cM(2,5)$ has $k=-4/3$.

The central charge of the minimal models is given by $c(p',p) = 1-\frac{6(p'-p)^2}{pp'}$. The primary fields of the minimal models are labeled by integers $(r,s)$, which take the values $1\leq r < p'$ and $1\leq s < p$. The labels $(r,s)$ and $(p'-r,p-s)$ correspond to the same primary field. Finally, the conformal dimensions of the fields are given by
\begin{equation}
h(r,s) = \frac{(rp-sp')^2-(p'-p)^2}{4pp'} \ .
\end{equation}

\subsection{Gepner parafermions in terms of minimal models}

The non-Abelian part of the CFT describing the NASS state, was originally written in terms of Gepner parafermions, \cite{gepner1987} which can be expressed in terms of the coset,
\[
{\rm GPf}[su(3)_k]= \frac{su(3)_k}{u(1)_{2k} u(1)_{6k}}.
\]
For our present purposes, we shall focus on $k=2$, in which case the above coset is equivalently written as
\begin{equation}
{\rm GPf}[su(3)_2] = \frac{su(2)_1\times su(2)_1 \times su(2)_1}{su(2)_3}
\end{equation}
[for arbitrary $su(3)_k$ parafermions, $su(2)$ has to be replaced with $su(k)$ in the above]. If one multiplies the numerator and denominator of the coset above by $su(2)_2$, one can factorize the coset into the (semidirect) product of the minimal models
$\cM(3,4)$ and $\cM(4,5)$,
\begin{align}
{\rm GPf}[su(3)_2] &\approx \frac{su(2)_1\times su(2)_1 \times su(2)_1\times su(2)_2}
{su(2)_2 \times su(2)_3}  \nonumber \\ &\approx \cM(3,4) \ltimes \cM(4,5).
\end{align}

Let us be more precise about this correspondence. The {\em direct} product
$\cM(3,4) \times \cM(4,5)$ does not correspond to the $su(3)_2$ Gepner parafermions. Instead, one has to consider
a so-called different modular invariant, \cite{cappelli1987a,cappelli1987b} constructed from the fields present in the direct product.
To do that, we will follow the logic presented in Refs.~\onlinecite{bais2009,fuchs1996,frohlich2004}. This amounts to identifying a boson in the CFT, which is then said to be ``condensed'' (or added to the chiral algebra). To identify a suitable boson, we give the fields of the models
$\cM(3,4)$ and $\cM(4,5)$ in Table~\ref{Tab:KacM4534}. (Fusion rules for the minimal models are given in Refs. \onlinecite{byb,belavin1984}.) 

\begin{table}[ht]
  \begin{tabular}{r|ccc}
  \hline \hline
  \multicolumn{4}{c}{$\mathcal{M}(3,4)$} \\
  \hline 
  $h(r,s)$ & $s=1$ & $2$ & $3$\\
  \hline
  $r=1$ & 0 & 1/16 & 1/2 \\
  $2$     & 1/2 & 1/16 & 0 \\
  \hline \hline
  \end{tabular}
  \hskip 8mm
  \begin{tabular}{r|cccc}
  \hline \hline
  \multicolumn{5}{c}{$\mathcal{M}(4,5)$} \\
  \hline 
  $h(r,s)$ & $s=1$ & $2$ & $3$ & $4$\\
  \hline
  $r=1$ & $0$ & $1/10$ & $3/5$ & $3/2$\\
  $2$ & $7/16$ & $3/80$ & $3/80$ & $7/16$\\ 
  $3$ & $3/2$ & $3/5$ & $1/10$ & $0$\\
   \hline \hline
  \end{tabular}
   \caption{Kac table of conformal weights for the unitary minimal models $\mathcal{M}(3,4)$ and
   $\mathcal{M}(4,5)$.}
  \label{Tab:KacM4534}
\end{table}

We will label the fields in the product theory in terms of the conformal dimensions (or conformal weights) of the contributing fields of the original CFTs. We find that there is indeed a Bosonic field (i.e., a field with integer scaling dimension) in the product, namely $(1/2,3/2)$. The condensation picture amounts to the following procedure: \cite{bais2009} Particles which can be obtained from one another by fusion of the boson, are ``identified'' (in the same way as the boson itself is identified with the vacuum, or trivial particle); in addition, particles which are not mutually local with the boson, are ``confined''; a particle which is mutual local with the boson does not generate a phase factor when transported around the boson; finally, it can happen that particles ``split''. We will see an example of splitting below. In this section, we will be rather brief; more details on how one constructs the correct theory can be found in the Appendix.

To get started, we list the particles in the product theory which are mutually local with this boson, namely
\begin{align*}
e &= (0,0) = (1/2,3/2),\\
(1/10)_a &= (1/16,3/80),\\
(1/10)_b &= (0,1/10) = (1/2,3/5),\\
(1/2)_a &= (1/16,7/16),\\
(1/2)_b &= (0,3/2) = (1/2,0),\\
3/5 &= (0,3/5) = (1/2,1/10).
\end{align*}
The equalities within one line signify that the particles corresponding to the two fields are identified. As an example, one obtains $(1/2,3/5)$ from
$(0,1/10)$ by fusion with the boson, namely $(0,1/10) \times (1/2,3/2) = (1/2,3/5)$.
We also note that the particles $(\frac{1}{2})_a$ and $(\frac{1}{10})_a$ have to split into two particles, thereby giving the full $Z_3$ symmetry present in the $su(3)_2$ parafermion theory. That this splitting is necessary can be seen by taking, for instance, the fusion product $(\frac{1}{2})_a\times (\frac{1}{2})_a = (0,0) + (1/2,0) + (0,3/2) + (1/2,3/2)$. Because both $(0,0)$ and $(1/2,3/2)$ correspond to the identity, splitting of the field $(\frac{1}{2})_a$ is necessary. A similar argument applies for $(\frac{1}{10})_a$, and one can convince oneself that in the end, one indeed obtains the correct fusion rules of the $su(3)_2$ parafermion theory. We conclude this section by mentioning that the example of the equivalence
${\rm GPf} [su(3)_2] \approx \cM(3,4) \ltimes \cM(4,5)$ featured prominently in a paper relating the Moore--Read and NASS states.\cite{grosfeld2009}

\subsection{CFT for the spin-singlet Gaffnian}

The spin-singlet Gaffnian that we aim to construct should have the same vanishing properties as the Gaffnian when considered as a function of only one spin species of particles. Motivated by the product of minimal models describing the $su(3)_2$ Gepner parafermions, we take as a starting point the coset
\begin{equation*}
\frac{su(2)_1 \times su(2)_1 \times su(2)_{k} }{su(2)_{k+2}} \ ,
\end{equation*}
with $k$ some fraction. Multiplying numerator and denominator by $su(2)_{k+1}$, we can write
\begin{equation*}
\frac{su(2)_1 \times su(2)_{k} }{su(2)_{k+1}} 
\times
\frac{su(2)_1 \times su(2)_{k+1} }{su(2)_{k+2}}  .
\end{equation*}
The CFT describing the Gaffnian is the nonunitary minimal model $\cM(3,5)$. To retain $\cM(3,5)$ in the above construction,  we will choose $k=-1/2$. With this choice, our coset takes the form $\cM(3,5)\times \cM(5,7)$. In analogy with the Gepner parafermion case above, we look for a field with integer scaling dimension in the product theory, and condense it. The field content of the models is given in Table~\ref{Tab:KacM5735}.  (Fusion rules for the minimal models are given in Refs. \onlinecite{byb,belavin1984}.) 

\begin{table}[ht]
  \begin{tabular}{r|cccc}
  \hline \hline
  \multicolumn{5}{c}{$\mathcal{M}(3,5)$} \\
  \hline 
  $h(r,s)$ & $s=1$ & $2$ & $3$ & $4$\\
  \hline
  $r=1$ & 0 & $-$1/20 & 1/5 & 3/4 \\
  $2$     & 3/4 & 1/5 & $-$1/20 & 0 \\
  \hline \hline
  \end{tabular}
  \vskip 8mm
  \begin{tabular}{r|cccccc}
  \hline \hline
  \multicolumn{7}{c}{$\mathcal{M}(5,7)$} \\
  \hline 
  $h(r,s)$ & $s=1$ & $2$ & $3$ & $4$ & $5$ & $6$\\
  \hline
  $r=1$ & 0 & 1/28 & 3/7 & 33/28 & 16/7 & 15/4 \\
  $2$ & 11/20 & 3/35 & $-$3/140 & 8/35 & 117/140 & 9/5 \\ 
  $3$ & 9/5 & 117/140 & 8/35 & $-$3/140 & 3/35 & 11/20 \\
  $4$ & 15/4 & 16/7 & 33/28 & 3/7 & 1/28 & 0 \\
   \hline \hline
  \end{tabular}
   \caption{Kac table of conformal weights for the unitary minimal models $\mathcal{M}(3,5)$ and
   $\mathcal{M}(5,7)$.}
  \label{Tab:KacM5735}
\end{table}

It turns out that the only boson that we can condense (or add to the chiral algebra), is the field $(9/5,1/5)$, which curiously has nontrivial fusion rules. The only fields that survive this condensation are found to correspond to product of fields of the first column of the Kac table for $\cM(3,5)$ and the fields of the first row of the Kac table for $\cM(5,7)$. In addition, none of these product fields are split. It follows that the resulting theory can be thought of as the product of a nonunitary semion theory (two fields with $Z_2$ fusion rules and dimensions $0$ and $3/4$) and a theory of six fields, satisfying $su(2)_5$ fusion rules. This second theory is also nonunitary, and the scaling dimensions are again simply read off from the Kac table, $(0,1/28,3/7,33/28,16/7,15/4)$. We note that this theory is modular (inherited via the coset construction). Here, we were rather brief in our description of the construction of the theory $\mathcal{M}(3,5) \ltimes \mathcal{M}(5,7)$. In Sec. 2 of the Appendix, we will give the details of the construction.

Before we start with the construction of the electron and quasihole operators, in Table~\ref{Tab:spingaffcft} we give the Kac table of the fields present in the CFT we constructed, 
$\mathcal{M}(3,5) \ltimes \mathcal{M}(5,7)$, which will constitute the non-Abelian part of the CFT describing the \wf{}. The full CFT also contains the $u(1)$ vertex operators (see, for instance, Ref.~\onlinecite{moore1991}). It is important to note that one has to be careful in determining the scaling dimensions of the fields in the product theory. The dimensions obtained by simply adding the scaling dimensions of the constituent fields can in fact correspond to the scaling dimensions of descendant fields, differing from the scaling dimensions of the primaries by integers. In particular, the field $(15/4,0)$ is identified with $(9/5,1/5) \times (15/4,0) = (11/20,1/5)$, which has scaling dimension $3/4$. Similarly, $(9/5,1/5) \times (15/4,3/4) = (11/20,-1/20)$, which has scaling dimension $1/2$. 

The fields occupying the corners of the Kac table, namely $(0,0), (0,3/4), (15/4,0), (15/4,3/4)$, are special, because they are simple currents. A simple current is a field, which when fused with any other field, always gives a single field as the result. Therefore, the particles corresponding to simple currents are Abelian. To construct the ``electron'' operator, one is therefore only allowed to use one of these four fields. Before we get started, we note that the Abelian sector of the theory at hand has less symmetry than the one describing the NASS state, which has three primary fields all with the same conformal weight. 

\begin{table}
\begin{tabular}{r|cccccc}
\hline
\hline
\multicolumn{7}{c}{$\mathcal{M}(3,5) \ltimes \mathcal{M}(5,7)$} \\
\hline 
$h(i,j)$ & $j=0$ & $1$ & $2$ & $3$ & $4$ & $5$\\
\hline
$i=0$ & 0 & 1/28 & 3/7 & 5/28 & 2/7 & 3/4 \\
$1$ & 3/4 & 11/14 & 5/28 & -1/14 & 1/28 & 1/2\\
\hline
\hline
\end{tabular}
\caption{Kac table of conformal weights
of the CFT describing the non-Abelian part of the spin Gaffnian.}
\label{Tab:spingaffcft}
\end{table}

We will now work with the assumption that we do have a $Z_2$ symmetry between the fields $(15/4,0)$ and $(0,3/4)$, [i.e., the fields with labels  $(i,j)=(0,5),(1,0)$] which we need, if the correlator is to describe a spin-singlet state (in the Appendix, we will give evidence supporting this statement). In addition, we want the wave function to look like a Gaffnian when viewed as a wave function for either spin-up or spin-down particles alone. We will introduce the following notation for the fields:
\begin{align}
\id &= (0,0), & \psi_\ua &= (0,3/4), \nonumber \\ \psi_\da &= (15/4,0), & \psi_{\ua\da} &= (15/4,3/4).
\end{align}
The fusion rules of these fields read $\psi_\ua \times \psi_\ua = \psi_\da \times \psi_\da = \psi_{\ua\da} \times \psi_{\ua\da} = \id$ and $\psi_\ua \times \psi_\da = \psi_{\ua\da}$.

We will continue with adding the appropriate vertex operators, giving charge and spin to the constituent particles, which is the usual procedure (see, for instance, Ref. ~\onlinecite{moore1991}). The guiding principle will be to construct electron operators, which give rise to the vanishing properties we want. The ansatz for the operators is
\begin{align}
V_{\ua} (z_\ua) &= \psi_\ua (z_\ua) e^{i \alpha \phi_c + i \beta \phi_s} (z_\ua) \nonumber, \\
V_{\da} (z_\da) &= \psi_\da (z_\da) e^{i \alpha \phi_c - i \beta \phi_s} (z_\da),
\end{align}
where $\alpha$ and $\beta$ are constants to be determined. The fields $\phi_c$ and $\phi_s$ are $u(1)$ compactified bosons. As we already pointed out, when we bring several up-particles together, we demand that the wave function behaves in the same way as the Gaffnian wave function. The electron operator in the CFT description of the Gaffnian reads $V (z) = \psi(z) e^{i \sqrt{\frac{3}{2}} \phi}(z)$, with $\psi(z)$ a field with conformal dimension $3/4$. Equivalences with this polarized case gives us the constraint $\alpha^2 + \beta^2 = 3/2$.

Moreover, bringing a spin-up and a spin-down particle together should not lead to a divergence. Using the fusion rule $V_\ua \times V_\da = \psi_{\ua\da} e^{2 i \alpha \phi_c}$ gives rise to a factor $(z_\ua-z_\da)^{-1 + \alpha^2-\beta^2}$ in the operator product expansion. To avoid a pole, one obtains the constraint $\alpha^2-\beta^2 = p$, with $p$ an integer greater then or equal to 1. Picking the minimal choice $p=1$, one finds that $\alpha = \sqrt{5/4}$ and $\beta = 1/2$. This value for $\beta$ is in fact equal to the value this parameter takes in the case of the NASS state.

With the electron operators
$V_{\ua} = \psi_\ua e^{i \sqrt{5/4}\phi_c + 1/2 i \phi_s}$
and
$V_{\da} = \psi_\da e^{i \sqrt{5/4}\phi_c - 1/2 i \phi_s}$
in place, we conclude that the corresponding wave function does not vanish when any two particles come together. It vanishes as three powers when three up (or three down) particles come together. In addition, when two up and one down particle come together, the wave function generically vanishes quadratically. We then assume that one can use logic similar to the NASS case, and thus, exploiting the SU(2) symmetry, \cite{ardonne2001} one finds that the wave function vanishes as three powers when three particles in the $S=3/2$ channel are coincident, and quadratically when three particles in the $S=1/2$ channel are coincident.

The filling fraction of the wave function is given by $\nu = 4/5$. The scaling dimension $h_e = 3/4 + 3/4 = 3/2$ gives rise to a shift $\delta = 3$, the same as for the Gaffnian. We will defer the construction of the quasihole operators to the Appendix, but mention here that they are constrained by the fact that the wave functions for the electrons should not have poles, even in the presence of quasiholes. For $N$ particles the ground-state \wf{} wave function can be written as a conformal block containing equal numbers of spin-up and spin-down electron fields:
\[
\Psi_{\mbox{\wf{}}} =\left \langle \psi_{\uparrow} (z^{\uparrow}_1) \ldots  \psi_{\uparrow} (z^{\uparrow}_{\frac{N}{2}})  \psi_{\downarrow} (z^{\downarrow}_{\frac{N}{2}+1}) \ldots  \psi_{\uparrow} (z^{\downarrow}_{N}) \right \rangle.
\]

Finally, we would like to mention that if we would have assumed that the field $\psi_{\ua,\da}$ would have had conformal dimension $3/2$ instead of $1/2$ (which is not consistent, because of the field identification above), one would have found $\alpha = \beta = \sqrt{3/4}$. The resulting wave function would then factorize as Gaffnian(up) $\times$ Gaffnian(down). The CFT description for such a wave function should be $\cM(3,5)\times \cM(3,5)$, which is at odds with the CFT description we have used here. 


\section{Ground-state wave function}
\label{secwave function}

In this section we shall present an explicit construction for the ground-state \wf{} wave function for the Bosonic case. Our construction is conjectured with the view to satisfying the vanishing properties as well as the constraints of filling factor $\nu=4/5$ and shift $\delta=3$, arising from the CFT considerations put forward in Sec.~\ref{secCft}. In addition, the wave function must describe a spin-singlet state. That condition specifies the requirement to satisfy the Fock cyclic symmetry conditions (see Ref. \onlinecite{ham}).

Before we describe how to construct the \wf{}, it will be useful to motivate our methodology by briefly reviewing the construction of the spin-polarized Gaffnian wave function. \cite{simon2007b} For the Gaffnian, the CFT requirements imposed on the vanishing properties are that for any three particles coincident the wave function must vanish as three powers and that the wave function must not vanish for any two particles coincident [the CFT describing the Gaffnian is the nonunitary minimal model $\mathcal{M}(3,5)$]. Further, it was determined that in its Bosonic form, the Gaffnian occurs at filling $2/3$ and $\delta=3$.

To give the expression of the Gaffnian wave function that can be generalized to the \wf{}, we divide the particles into two groups $A$ and $B$ of equal size. The Gaffnian wave function can then be written as
\begin{widetext}
\[
\Psi_{\mbox{Gaffnian}} = \hat S \left [
\prod_{i<j \in A}(z^{A}_i -z^{A}_j)^2  
\prod_{i<j \in B}(z^{B}_i -z^{B}_j)^2  
\prod_{i \in A, j \in B} (z^{A}_i -z^{B}_j) \;
\mathbf{Per} \left [ \frac{1}{z^{A}_i-z^{B}_j} \right ] 
\right ],
\]
\end{widetext}
where $\hat S$ represents a symmetrization operation over all $N$ particle coordinates, and
$\mathbf{Per} [ M_{ij}] $ denotes the ``permanent'' of a matrix $M$ whose elements are in this case given by $M_{ij} = (z^{A}_i-z^{B}_j)^{-1}$. On its own, this permanent factor contains a certain pattern of poles.
When placed within the full construction, these poles conspire to ensure that the overall wave function does not vanish as two particles become coincident. Thus the polynomial vanishes only when three particles are coincident. The vanishing power can be tuned by adjusting the exponents of each Jastrow-type factor in the construction. For example,  $\prod_{i<j \in A}(z^{A}_i -z^{A}_j)^2$ could be adjusted to, say, $\prod_{i<j \in A}(z^{A}_i -z^{A}_j)^4$ (and similar for the particles in group $B$) to alter the vanishing power from 3 to 5 in this example. An entirely nontrivial step is to determine whether or not these vanishing properties are retained once the overall symmetrization operation has been completed. For the Gaffnian, it was found that the vanishing properties are retained. 

Now let us turn to the construction of the ground-state wave function for the \wf{}. We have determined that the following construction, given in Eq.~(\ref{eqwave function}), gives a wave function at filling factor $\nu=4/5$ and $\delta=3$ and it satisfies the CFT vanishing constraints for $N=4,8,12$. We have reason to believe that it will work for all other values of $N$ (i.e., 16, 20, etc.), although an explicit check is not possible. Thus, we conclude that it must be proportional to the conformal block giving the \wf{} ground state. In the construction, we have used a clustering principle: The particle coordinates are first divided into equal sets of spin up and spin down, and then each of these sets is further subdivided into two equal subsets labelled by either $A$ or $B$, giving four sets in total ($A\uparrow$, $A\downarrow$, $B\uparrow$, $B\downarrow$). With the label $A$ or $B$ on its own, we refer to the two sub-sets with either spin direction,
\begin{widetext}
\begin{equation}
\label{eqwave function}
\Psi_{\mbox{\wf{}}} = \hat Y_{S=0}  \left [   \prod_{a=A,B} \left\{ \prod_{i<j \in a\uparrow}(z^{a\uparrow}_i -z^{a\uparrow}_j)^2   \prod_{i<j \in a\downarrow}(z^{a\downarrow}_i -z^{a\downarrow}_j)^2  \prod_{i \in a\uparrow, j \in a\downarrow}(z^{a\uparrow}_i -z^{a\downarrow}_j)  \right \}   \prod_{i \in A, j \in B} (z^{A}_i -z^{B}_j) \; \mathbf{Per} \left [ \frac{1}{z^{A}_i-z^{B}_j} \right ] \right ],
\end{equation}
\end{widetext}
where, as above, $\mathbf{Per} [ M_{ij}] $ denotes the permanent of a matrix $M$ whose elements are in this case given by $M_{ij} = (z^{A}_i-z^{B}_j)^{-1} $ and where $\hat Y_{S=0}$ is the Young operator for a spin-singlet representation of the symmetric group---this operation is required in order to guarantee that the wave function satisfies the correct Fock cyclic symmetry conditions for a spin-singlet state. \cite{ham}
\[
\hat Y_{S=0} = \hat S_{z^{\uparrow}_1 \ldots z^{\uparrow}_{N/2} } \hat S_{z^{\downarrow}_1 \ldots z^{\downarrow}_{N/2} } \hat A_{z^{\uparrow}_1 z^{\downarrow}_1}\ldots  \hat A_{z^{\uparrow}_{N/2} z^{\downarrow}_{N/2}},
\]
that is, the Young operator for a spin-singlet representation corresponds to the operation of antisymmetrizing over ordered pairs of spin-up and spin-down coordinates, followed by symmetrizating over all spin-down and then all spin-up coordinates. Crucially, it is important to note that the wave function does not completely vanish when $\hat Y_{S=0}$ is applied!

In Sec.~\ref{secNumerics} we provided strong evidence showing that the spectrum of the Hamiltonian for the \wf{} contains a unique zero energy state with $L=0$ and $S=0$, for precisely the values of particle number and flux corresponding to filling $\nu=4/5$ and $\delta=3$. Therefore we conclude that this trial ground-state wave function is unique, and further, it is precisely the highest density ground state of the \wf{} Hamiltonian proposed in Eq.~(\ref{eqHamiltonian}).


\section{Discussion}
\label{discussion}

To summarize our findings, we have presented evidence to demonstrate that the proposed \wf{} wave function satisfies many of the ``ingredients of an ideal theory of a FQHE state'': A local Hamiltonian, a relatively simple analytic expression for the wave function, at least for the ground state, and a correspondence of that wave function to a 2D rational CFT. We have verified the quasihole spectrum of the \wf{} Hamiltonian, checking against both the spin dressed squeezing algorithm and the spinful version of the generalized Pauli principle. Indeed, this study is an interesting test case for the application of such techniques.

Concerning quasiholes, although we have made progress on the CFT description of the quasihole operators (see the Appendix), we have not yet determined simple, analytic expressions for the corresponding quasihole wave functions. For the spin-polarized Gaffnian, the quasihole wave functions have been constructed; \cite{simon2007b} the main stumbling block here is the additional complexity due to the spin degree of freedom.

In Sec.~\ref{secMethods}  we described how to generate the \wf{} wave functions by means of a local Hamiltonian written in terms of spin-dependent pseudopotentials. An interesting question left unaddressed is whether other states could be constructed with faster vanishing properties than the \wf{}. For instance, in the spin-polarized case, adding the next highest $L$ pseudopotential to the Gaffnian Hamiltonian is known to produce the Haffnian Hamiltonian. \cite{green2002} Might we be able to generate a ``spin-Haffnian''  state with a Hamiltonian containing positive
$ V^3_{0,3/2}$, $ V^3_{2,3/2,}$, $ V^3_{1,1/2}$ and now, in addition, $V^3_{3,3/2}$ (and possibly $ V^3_{2,1/2}$)? Wave functions of this type were also considered in Ref.~\onlinecite{estienne2011}.
Presently we lack a corresponding CFT description with which to conduct the same checks as for the \wf{}. It is also unclear at the present time whether that Hilbert space could identically be constructed by a squeezing algorithm or generalized Pauli principle approach. 

Concerning CFT coset constructions, it is noteworthy that we were able to derive a self-consistent CFT from the product of two nonunitary minimal models by constructing a nondiagonal modular invariant. Specifically, the nonunitary models in the product can be thought of as cosets of fractional level affine Lie algebras. The resulting CFT corresponds to the nonunitary coset $su(2)_1\times su(2)_1 \times su(2)_{k-2}/su(2)_k$, with $k$ fractional. In Sec.~2 of the Appendix, we deal with a more general case.

In general, for \emph{any} unitary coset construction with simple current extensions there is a proven procedure to generate the CFT. \cite{fuchs1996,frohlich2004} For nonunitary cosets, no such general procedure exists, and each CFT must be constructed on a case-by-case basis. \cite{byb} In the CFT construction presented in this work, we have provided a further example of a case where the construction of non-diagonal modular invariants for nonunitary CFTs is possible.

Another powerful tool for the analysis of the quantum Hall wave functions has been the concept of Jack polynomials. \cite{bernevig2008b,estienne2011} The Jack polynomials provide a convenient basis in which to describe polynomial wave functions with precisely defined vanishing properties.  Based on this new insight, it has been determined that states such as that of Moore and Read---where there exists a corresponding local Hamiltonian, a relatively simple analytic form for the wave function, and a CFT description of the state---are in fact simply special cases within a much broader classification of FQHE states in terms of these Jack polynomials. It has been further realized that such states are rather atypical: Out of all the Jacks, there exist only a handful of special cases for which all three of these ideal ingredients of a theory of a FQHE state are present. In this paper we have presented a further example, albeit that the CFT corresponding to the SSG is nonunitary implying a gapless, compressible wave function. 

\acknowledgements

We thank J.~K.~Slingerland, B.~Schellekens, B.~Estienne and B.~A.~Bernevig for helpful discussions. We also thank B.~Estienne for useful comments on the manuscript. S.C.D. and S.H.S. were supported by EPSRC Grants No. EP/I032487/1 and No. EP/I031014/1. N.R. was supported by  NSF CAREER DMR-095242, ONR-N00014-11-1-0635,  ARMY-245-6778, MURI-130-6082, ANR-12-BS04-0002-02, Packard Foundation, and a Keck grant. We thank NORDITA, the Institute Henri Poincar\'e and the Aspen Centre for Physics for hospitality during our collaboration.


\appendix

\section*{Appendix: More Details about the CFT}
\label{appendixCft}

\subsection{Quasihole operators}

In this Appendix, we will construct the operators corresponding to the quasiholes with the smallest quantum numbers. From them, we construct the set of all quasihole operators, whose number gives the torus degeneracy.

We start with the field in the CFT $\cM(3,5) \ltimes \cM(5,7)$, and add the $u(1)$ factors (as we did in the construction of the electron operators), in such a way that the quantum numbers are minimized. The fields in the $\cM(3,5)\ltimes \cM(5,7)$ theory are denoted by $\phi_{i,j}$, with $i=0,1$ and $j=0,1,2,3,4,5$. The label $i$ adds modulo $2$ under fusion, while the label $j$ obeys $su(2)_5$ fusion rules. For general $k$, the $su(2)_k$ fusion rules read 

$$j_1 \times j_2 = \sum_{j_3 = |j_1-j_2|}^{\min(j_1+j_2,k-j_1-j_2)} j_3 \ .$$

The two fields with the smallest scaling dimension are $\phi_{0,1}$ and $\phi_{1,4}$, so we start by writing the ansatz for the smallest quasihole operator as $V_{\rm qh} (w) = \phi_{0,1} e^{i a \phi_c + i b \phi_s}$. Fusing this quasihole with an electron should give a wave function which is still analytic in the electron coordinates. This leads to the constraints
$a \sqrt{5/4} + b/2 -1/2 = p$ and $a \sqrt{5/4} - b/2  = q$, where $p$ and $q$ are non-negative integers. Upon picking the minimal choice $p=q=0$, one finds $a = 1/(2\sqrt{5})$ and $b = 1/2$, which corresponds to a quasihole with charge $1/5$ and spin $s_z = 1/2$. The operator reads $V_{{\rm qh},\ua} = \phi_{0,1} e^{i\frac{1}{2\sqrt{5}}\phi_c + i/2 \phi_s}$.

Constructing the other minimal quasihole, based on $\phi_{1,4}$, one obtains $V_{{\rm qh},\da} = \phi_{1,4} e^{i\frac{1}{2\sqrt{5}}\phi_c - i/2 \phi_s}$, with charge $1/5$ and $s_z = -1/2$. Starting from these ``fundamental'' quasiholes, one obtains all the quasihole species by successive fusion.

When we construct the other quasiholes, we will do this ``modulo the electron operators'', because two particles which can be obtained from each other by fusion of an ``electron'' correspond to the same species of particle. Thus, we will first determine the operators which one can obtain by fusing (possibly different species of) electrons.

We first introduce some notation for the labels of the different fields (or species of particles). The fields will be denoted by their three ``quantum numbers'', namely the field of the non-Abelian CFT, the charge, and the spin. As an example, the ``electron'' operators are given by $(\phi_{0,5},1,\frac{1}{2})$ and $(\phi_{1,0},1,-\frac{1}{2})$. The ``composite'' of these two operators is $(\phi_{1,5},2,0)$. It will be useful to consider the ``dual'' operators, which read $(\phi_{0,5},-1,-\frac{1}{2})$, $(\phi_{1,0},-1,\frac{1}{2})$ and $(\phi_{1,5},-2,0)$. From these operators, we can construct particles without charge, for instance $(\phi_{0,5},1,\frac{1}{2})\times (\phi_{1,0},-1,\frac{1}{2}) = (\phi_{1,5},0,1)$ and $(\phi_{1,0},1,-\frac{1}{2})\times (\phi_{0,5},-1,-\frac{1}{2}) = (\phi_{1,5},0,-1)$. Since fusing a quasihole with any of the operators above does not give us a new species of quasihole, it follows that the set of independent quasiholes can be labeled such that the charge and spin take the values $0\leq q < 1$ and $s_z = 0, 1/2$, respectively.

The quasihole operators we constructed above can be written as $(\phi_{0,1},\frac{1}{5},\frac{1}{2})$ and $(\phi_{1,4},\frac{1}{5},-\frac{1}{2})$. These two fields are in fact to be identified, because $(\phi_{1,4},\frac{1}{5},-\frac{1}{2})\times (\phi_{1,5},0,1) = (\phi_{0,1},\frac{1}{5},\frac{1}{2})$. Thus, to construct all the different topological sectors, it suffices to repeatedly fuse the quasihole $(\phi_{0,1},\frac{1}{5},\frac{1}{2})$, and record the different sectors, modulo the ``electron'' operators.

This procedure leads to the following topological sectors:

\begin{center}
\begin{tabular}{ccccc}
& $(\phi_{0,5},\frac{1}{5},\frac{1}{2})$ && $(\phi_{1,0},\frac{3}{5},\frac{1}{2})$ & \\
$(\phi_{0,4},0,0)$ && $(\phi_{1,1},\frac{2}{5},0)$ && $(\phi_{0,4},\frac{4}{5},0)$ \\
& $(\phi_{0,3},\frac{1}{5},\frac{1}{2})$ && $(\phi_{1,2},\frac{3}{5},\frac{1}{2})$ & \\
$(\phi_{0,2},0,0)$ && $(\phi_{1,3},\frac{2}{5},0)$ && $(\phi_{0,2},\frac{4}{5},0)$ \\
& $(\phi_{0,1},\frac{1}{5},\frac{1}{2})$ && $(\phi_{1,4},\frac{3}{5},\frac{1}{2})$ & \\
$(\phi_{0,0},0,0)$ && $(\phi_{1,5},\frac{2}{5},0)$ && $(\phi_{0,0},\frac{4}{5},0)$. \\
\end{tabular}
\end{center}
The fields which would appear in the next column, with charge $1$, can be identified to the fields in the first column, with charge $0$. We thus find that the number of different topological sectors is $15$, which is the degeneracy on the torus.

It is interesting to note that these sectors can also be understood from the so-called ``thin-torus limit,''  \cite{bergholtz2005,bergholtz2006,seidel2006} the generalized Pauli principle, \cite{estienne2011} or the squeezing patterns we discussed in Sec.~\ref{secSqueezing}.

First, the pattern which we used to define the spin-singlet Gaffnian via squeezing, is $(2,0,2,0,0)$, i.e., two neighbouring orbitals can host maximally two particles, and three through five consecutive orbitals can be occupied by maximally four particles. This gives rise to the sectors $(2,0,2,0,0,2,0,2,0,0)$ (five times on a torus), $(2,0,1,1,0,2,0,1,1,0)$ (five times on a torus), and $(1,1,1,1,0,1,1,1,1,0)$ (five times on a torus). Each of the fields $(\phi_{0,0},0,0)$, etc., corresponds to one pattern. In the first column, we find $(\phi_{0,0},0,0) \equiv (2,0,2,0,0)$, $(\phi_{0,2},0,0) \equiv (1,0,1,1,1)$ and $(\phi_{0,4},0,0) \equiv (0,2,0,1,1)$. The other fields can be obtained by fusing with $(\phi_{0,1},\frac{1}{5},\frac{1}{2})$, which brings one from one column to the adjacent one on the right. On the level of the patterns, one has to to ``hop'' one particle one step to the right, assuming periodic boundary conditions. Thus, one finds $(2,0,2,0,0) \rightarrow (2,0,1,1,0)$, while in the other two cases, there are two options,  $(1,0,1,1,1)\rightarrow (2,0,1,1,0);(0,1,1,1,1)$ and $(0,2,0,1,1)\rightarrow (0,1,1,1,1);(0,2,0,0,2)$. Here, we have the identification $(\phi_{0,1},\frac{1}{5},\frac{1}{2}) \equiv (2,0,1,1,0)$, $(\phi_{0,3},\frac{1}{5},\frac{1}{2}) \equiv (0,1,1,1,1)$ and $(\phi_{0,5},\frac{1}{5},\frac{1}{2}) \equiv (0,2,0,0,2)$. The remaining columns follow in the same way.

\subsection{Field content and the modular invariant of the product CFT}
\label{app:field-content}

To completely specify the conformal field theory describing the spin-singlet Gaffnian, we need to give the field content of the product CFT we use. In Sec.~\ref{secCft} above, we used the condensation picture to argue which fields are present in the final CFT we used. We did not in fact go through the condensation picture in full detail: In particular, we glossed over the fact that the boson itself has to split (because the fusion of the boson with itself has multiple fusion channels, including the vacuum and the boson itself). So, in the construction of the \wf{} CFT it is only one part of the boson which condenses, and going through the whole condensation procedure can become cumbersome.

In this Appendix, we will construct a modular invariant partition function of the product theory, using the boson condensation as a guiding principle. The first step of the procedure is to find a boson which can be condensed and to find the fields local with respect to that boson. Such fields will be present in the final CFT. Upon acting with the boson on these fields, we obtain the fields (in the original product theory) which are identified with those fields which are local with respect to the boson. In this way, we can construct the modular invariant partition functions. We refer to Refs.~\onlinecite{cappelli1987a,cappelli1987b,bais2009} for more details.

We start with those CFTs corresponding to wave functions with $n=2$ components, i.e., the generalizations of the NASS state, where the condensate particles have spin $1/2$. The CFT describing these states consists of a product of two minimal models, which, for $r>2$, are nonunitary (the case $r=2$ is the NASS state).

We already argued that for $r=3$, the starting point is the product theory $\cM (3,5) \times \cM(5,7)$. For general $r$, we instead find that we need $\cM (3,r+2) \times \cM(r+2,2r+1)$ as the starting product theory. We note that for the minimal model theory $\cM(p',p)$ to be well defined, $p'$ and $p$ have to be co-prime, which means that $r \bmod 3 \neq 1$.  We will assume that this relation is satisfied from now on.

The fields of the product theory will be labeled by $(r_1,s_1;r_2,s_2)$, with the usual field identifications, coming from the field identifications present in the minimal model $\cM(p',p)$. In the product theory, the field $b=(3,1;1,3)$ always has scaling dimension $h_b =2$, and hence is a boson. For $r=2$, we find $b \times b = (1,1;1,1)$, the identity. For $r=3$, we have $b \times b = (1,1;1,1)+(1,1;1,3)+(3,1;1,1)+(3,1;1,3)$. For $r \geq 5$ (but  $r \bmod 3 \neq 1$), we find the general result $b \times b = (1,1;1,1)+(1,1;1,3)+(1,1;1,5)+ (3,1;1,1)+(3,1;1,3)+(3,1;1,5)+(5,1;1,1)+(5,1;1,3)+(5,1;1,5)$.

We now describe the field content of the CFTs we employ in this paper, by specifying the relevant modular invariant partition functions. In general, these are given in terms of the characters of the holomorphic and antiholomoriphic part of the CFT. We will denote these characters by $\chi_r$ and $\overline{\chi}_{r'}$, respectively, where both $r$ are $r'$ are labels of the fields in the product CFT. The putative partition functions take the form $Z(\tau) = \sum_{r,r'} M_{r,r'} \chi_r (\tau) \overline{\chi}_{r'} (\overline{\tau})$, where the $M_{r,r'}$ are non-negative integers denoting the multiplicities of the holomorphic-antiholomorphic combination $\chi_r (\tau) \overline{\chi}_{r'} (\overline{\tau})$. The identity field should be present and nondegenerate, $M_{1,1} = 1$. In order that the partition function is modular invariant, the matrix $M$ should commute with both $S$ and $T$, the generators of modular transformations. The commutation condition can be seen to enforce that the only combinations $\chi \overline{\chi}$ which can occur have $h-\overline{h} \bmod 1 = 0$.

The so-called diagonal invariant, $M_{r,r'} = \delta_{r,r'}$ always exists, but this is not the invariant we are interested in here. We are interested in the invariant, which contains $4r$ fields, which are ``composed'' out of several fields, which are identified via the condensation of the boson present in the original product theory. To specify the invariants, we have to distinguish two cases separately. We start with $r$ odd. In this case, the fields in the final theory can be labeled by $(1,s_1;r_2,1)$, for $s_1=1,2$, and $r_2 = 1,\ldots,2r$, for a total of $4r$ fields. These fields are identified with the fields $(q,s_1,r_2,q)$, with $q=1,3,\ldots,r$, i.e., all possible odd values of the labels $r_1$ and $s_2$. In particular, the modular invariant partition function takes the form
\begin{widetext}
\begin{equation}
Z(\tau) =
\sum_{r_2=1}^{2r} \bigl| \chi_{(1,1;r_2,1)} + \chi_{(3,1;r_2,3)} + \cdots + \chi_{(r,1;r_2,r)} \bigr|^2 + 
\sum_{r_2=1}^{2r} \bigl| \chi_{(1,2;r_2,1)} + \chi_{(3,2;r_2,3)} + \cdots + \chi_{(r,2;r_2,r)} \bigr|^2  \ .
\end{equation}
In the case that $r$ is even, the new theory still contains the fields which are labeled by $(1,1;r_2,1)$, but the fields with labels $(1,2;r_2,1)$ are absent. Instead, the fields with labels $(2,2;r_2,2)$ are now present. These are identified with the fields $(q,2;r_2,q)$, where $q = 2, 4,\ldots, r$ is even. The partition function reads
\begin{equation}
Z(\tau) =
\sum_{r_2=1}^{2r} \bigl| \chi_{(1,1;r_2,1)} + \chi_{(3,1;r_2,3)} + \cdots + \chi_{(r+1,1;r_2,r+1)} \bigr|^2 + 
\sum_{r_2=1}^{2r} \bigl| \chi_{(2,2;r_2,2)} + \chi_{(4,2;r_2,4)} + \cdots + \chi_{(r,2;r_2,r)} \bigr|^2  \ .
\end{equation}
\end{widetext}

We will now consider the case where the particles have $n$ internal states rather than 2 internal states as they do for spin $1/2$. Let us focus on the case $n=3$, which would correspond to three-component states, such as the spin-1 states considered in Refs.~\onlinecite{reijnders2002,reijnders2004} . The generalization to higher $n$ will be clear after that. For $n=3$, the starting product theory is $\cM(3,r+2)\times\cM(r+2,2r+1)\times\cM(2r+1,3r)$. The Bosonic field which one can take as the field which condenses is $b=(3,1;3,3;1,3)$, which has scaling dimension $h_b = 2$. As was the case for $n=2$, we need that $r \bmod 3 \neq 1$, such that the factors in the product theory are well defined.

Starting with $r$ odd again, we can label the fields of the new (block-diagonal) modular invariant as follows. There are a few different groups, namely $(1,i;1,1;j,1)$, with $i=1,2$ and $j=1,2,\ldots,3r-1$, in addition to $(1,i;2,1;j,2)$, also with $i=1,2$ and $j=1,2,\ldots,3r-1$. The fields appearing in the blocks labeled by $(1,i;1,1;j,1)$ are $(q_1,i;q_2,q_1;j,q_2)$, where $q_1,q_2 \bmod 2 = 1$. In the case of the blocks labeled by $(1,i;2,1;j,2)$, the fields appearing in these blocks are $(q_1,i;q_2,q_1;j,q_2)$, but now with $q_1 \bmod 2 = 1$ and $q_2 \bmod 2 = 0$.

In the case that $r$ is even, the labels of the blocks are slightly different, in the same way as was the case for $r=2$. There are for different types of labels, namely $(1,1;1,1;j,1)$, $(1,1;2,1;j,2)$, $(2,2;1,2;j,1)$, and $(2,2;2,2;j,2)$, all with $j=1,2,\ldots,3r-1$. The fields appearing in these blocks are of the form $(q_1,i;q_2,q_1;j,q_2)$, where $q_1$ and $q_2$ are either even or odd, depending on the type of block they belong to. In either case $r$ even or odd, the number of fields in the block-diagonal modular invariant is given by $4 (3r-1)$.

A few remarks about the case of general $n$. The product theory one starts with is $\cM(3,r+2)\times\cM(r+2,2r+1)\times\cdots\times\cM(3+(n-1)(r-1),3+n(r-1))$, where $r\bmod 3 \neq 1$. The boson with scaling dimension $h_b=2$ has the labels $(3,1;3,3;\ldots;3,3;1,3)$. The number of different blocks, i.e., the number of fields in the new modular invariant, is $2^{n-1} [2+n(r-1)]$. The precise form of the labels of these blocks, and the fields appearing in them, depends on $r$ being even or odd, in a way which should be clear from the case $n=3$ above. The general form of the fields in the blocks is given by $(q_1,i;q_2,q_1;q_3,q_2;\ldots;q_{n-1},q_{n-2};j,q_{n-1})$, where $i=1,2$, $j=1,2,\ldots, [2+n(r-1)]$, and $q_2,q_3,\ldots,q_{n-1}$ can either be even or odd. For $r$ odd, $q_1$ is always odd, while for $r$ even we find that $i$ and $q_1$ have the same parity, $(i+q_1) \bmod 2 = 0$.

The invariants we discussed above are relevant for paired states with $k=2$, based on particles with spin $(n-1)/2$, with the property that the wave function vanishes as $r$ powers when three particles with the same $s_z$ come together. It is possible to generalize this construction to arbitrary clustered states with $k>2$, but these will be based on cosets of the type $su(k)_1 \times su(k)_1/ su(k)_2$, etc., making the structure of the modular invariants somewhat more involved. We will not deal with the case $k>2$ here.


\end{document}